\documentclass[seceq]{ptptex}

\usepackage{graphicx}
\usepackage{wrapft}

\title{
Super-elastic collisions in a thermally activated system%
}

\author{
Hiroto \textsc{Kuninaka} and Hisao \textsc{Hayakawa}
}

\inst{
Dept. of Physics, Chuo Univ.\\
Yukawa Institute of Theoretical Physics
}

\abst{
Impact phenomena of small clusters subject to thermal fluctuations 
 are numerically investigated. 
 From the molecular dynamics simulation for colliding two identical 
 clusters,  
 it is found that 
 the restitution coefficient for head-on collisions can exceed unity  
 when the colliding speed is smaller than the thermal velocity of 
 one ``atom'' of the clusters.  The averaged behavior can be understood by 
 the quasi-static theory of impact phenomena. 
 It is also confirmed that our result is governed by the fluctuation theorem.
}

\begin{document}
\maketitle
\section{Introduction}\label{intro}
Inelastic collisions are the process that a part of the initial kinetic energy 
of colliding bodies is distributed into the internal degrees of freedom. 
Although it is believed that the relative speed of colliding two bodies 
decreases after a collision, 
 the event that the relative rebound speed becomes larger 
 than the relative colliding speed is not contradicted
 with the energy conservation law. 
 Such an anomalous impact is prohibited by the second law of thermodynamics
 in which  the energy does not come back 
 to the macroscopic degrees of freedom
 once a part of energy is distributed into the microscopic ones.

For classical colliding bodies at the initial temperature $T$,   
the second law may be written as
\begin{equation}\label{eq1}
\frac{1}{2} \mu {\tilde V}^{2}+O(T)\ge \frac{1}{2} \mu {\tilde V}^{'2}
\end{equation}
where $\mu$ is the reduced mass of the colliding bodies, 
${\tilde V}$ and ${\tilde V}^{'}$ are respectively 
the relative colliding speed and the relative rebound speed.
~\cite{maes-tasaki} 
The second term in the left hand side is negligible 
in the case of macroscopic bodies. Thus, in the case of 
head-on collisions of macroscopic bodies, 
the restitution coefficient,  
$e \equiv {\tilde V}^{'}/{\tilde V}$, 
becomes less than unity~\cite{tasaki-jsp} while 
the restitution coefficient projected 
into the normal direction of the collision 
can exceed unity in the case of oblique collisions.
~\cite{louge,kuninaka_prl}  
The second term in the left hand side of eq.(~\ref{eq1}), however, 
is not negligible for
small systems. In this paper we investigate the statistical properties of
``super-elastic'' collisions for small colliding bodies 
where $e$ is larger than unity in head-on collisions.

``Super-elastic'' collisions  may be related to the fluctuation 
theorem~\cite{evans,evans_cohen_morris,gallavotti,crooks,PT_FT}  
in which the probability of the entropy production is related 
to the entropy absorption as 
$\frac{P(S^{\tau}=A)}{P(S^{\tau}=-A)} = \exp\left(\tau A\right)$,
where $P(S^{\tau}=A)$ is the probability distribution of observing 
the time averaged entropy production $S^{\tau}$ during time interval 
$\tau$ lies in the range $A$ to $A+dA$. 
Thus, it might be important to study the relation between 
the fluctuation theorem and ``super-elastic'' collisions 
due to large thermal fluctuations.  

There are numerical and theoretical studies for coalescence, 
scattering, and fragmentation of colliding nanoclusters. 
~\cite{kalweit, full, clus_col} Most of the low-speed collisions of 
nanoclusters cause coalescence.
However, some stable clusters such as fullerene can keep their form 
in collisions. ~\cite{full} 
Therefore, we focus on the properties of small stable clusters in which 
the interaction between two clusters is dominated 
by the repulsive force.

In this paper, we perform the molecular dynamics simulation of 
colliding repulsive clusters to investigate the effect of 
thermal fluctuations. The organization of this paper is as follows. 
In the next section, we introduce our model. In \S \ref{results}, 
 we investigate the relation between the restitution coefficient 
and colliding speed and that between the compressive force and 
the deformation. We also compare our numerical results 
with the fluctuation theorem.  
Section \ref{discussion} and \ref{conclusion} are devoted to 
the discussion and the conclusion of our results, respectively. 

\section{Model}\label{model}
\begin{wrapfigure}{l}{6.6cm}
\begin{center}
\includegraphics[width=.25\textwidth]{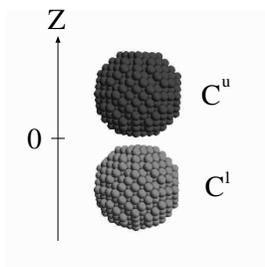}
\end{center}
\caption{Numerical model of colliding clusters. Each of them is composed of 
682 ``atoms'' which are bounded together by the Lennard-Jones potential.}
\label{fig1}
\end{wrapfigure}
Let us introduce our numerical model. Our model is composed of 
two identical clusters. Each of them is spherically cut from 
a face-centered cubic (FCC) lattice and consisted of $682$ ``atoms''. 
The clusters have facets due to the small number of ``atoms'' 
(Fig. ~\ref{fig1}). All the ``atoms'' in each cluster are bounded 
together by the Lennard-Jones potential $U(r_{ij})$ as 
\begin{equation}
U(r_{ij})=4\epsilon\left\{\left(\frac{\sigma}{r_{ij}}\right)^{12}-
               \left(\frac{\sigma}{r_{ij}}\right)^{6}\right\}, 
\end{equation}
where $r_{ij}$ is the distance between two ``atoms'', $i$ and $j$, in 
one cluster. $\epsilon$ is the energy constant and 
$\sigma$ is the lattice constant. 
When we regard the ``atom'' as argon, the values of the constants become 
$\epsilon=1.65\times10^{-21}\mathrm{J}$ and $\sigma=3.4$\AA, 
respectively.~\cite{rieth}
Henceforth, we label the upper and the lower clusters as 
cluster $C^{u}$ and cluster $C^{l}$, respectively. 
The interaction between the atom $k$ on the lower surface of $C^{u}$ 
and the atom  $l$ on the upper surface of $C^{l}$ is assumed to be 
the repulsive potential $R(r_{kl})=4\epsilon(\sigma/r_{kl})^{12}$, 
where $r_{kl}$ is the distance between the atoms $k$ and $l$. 
To reduce computational costs, we introduce the cut-off length $\sigma_{c}$ 
of the Lennard-Jones interaction as $\sigma_{c}=2.5 \sigma$. 

The procedure of our simulation is as follows. 
As the initial condition of simulation, 
the centers of mass of $C^{u}$  and $C^{l}$ are placed 
along the $z$-axis with the separation  $\sigma_{c}$ between them. 
The initial velocities of the ``atoms'' in both $C^{u}$ and $C^{l}$ 
obey Maxwell-Boltzmann distribution with the initial temperature $T$. 
The initial temperature is set to be $T=0.01 \epsilon$ or $T=0.02 \epsilon$ 
in most of our simulations.  
Sample average is taken over different sets of initial velocities 
governed by the Maxwell-Boltzmann velocity distribution for ``atoms''. 

To equilibrate the clusters, we adopt the velocity scaling 
method~\cite{haile,andersen} for 
$2000$ steps at the initial stage of simulations. 
We have checked the equilibration of the total energy 
in the initial relaxation process.
After the equilibration, 
we give translational velocities and the macroscopic rotations to $C^{u}$ and $C^{l}$ 
to make them collide against each other. 
The relative speed of impact 
ranges from ${\tilde V}=0.02 \sqrt{\epsilon/m}$ to 
${\tilde V}=0.07 \sqrt{\epsilon/m}$, 
which are less than the thermal velocity for one ``atom'' 
defined by $\sqrt{T/m}$, where $m$ is the mass of the ``atom''.

Numerical integration of the equation of motion for each atom 
is carried out by the second order symplectic integrator with 
the time step $dt=1.0 \times 10^{-2} \sigma/\sqrt{\epsilon/m}$. 
The rate of energy conservation, $|E(t)-E_{0}|/|E_{0}|$, 
is kept within $10^{-5}$, 
where $E_{0}$ is the initial energy of the system and $E(t)$ is the 
energy at time $t$.

We let the angle around $z-$axis, $\theta^{z}$, be $\theta^{z}=0$ 
when the two clusters are located mirror-symmetrically 
with respect to $z=0$. 
In most of our simulation, we set $\theta^{z}$ at $\theta^{z}=0$ 
as the initial condition. 
Let us comment on the dependence of 
the relative angle $\theta^{z}$ on the numerical results. 
From our impact simulation for 
$\theta^{z}_{i}=\pi i/18 \hspace{1mm} (i=1,...,9)$ at 
$T=0.02\epsilon$ we have confirmed that the initial orientation 
does not largely 
affect the restitution coefficient.


\section{Results}\label{results}
\begin{figure}[th]
\begin{center}
 \begin{minipage}{0.47\textwidth}
  \includegraphics[width=0.8\textwidth]{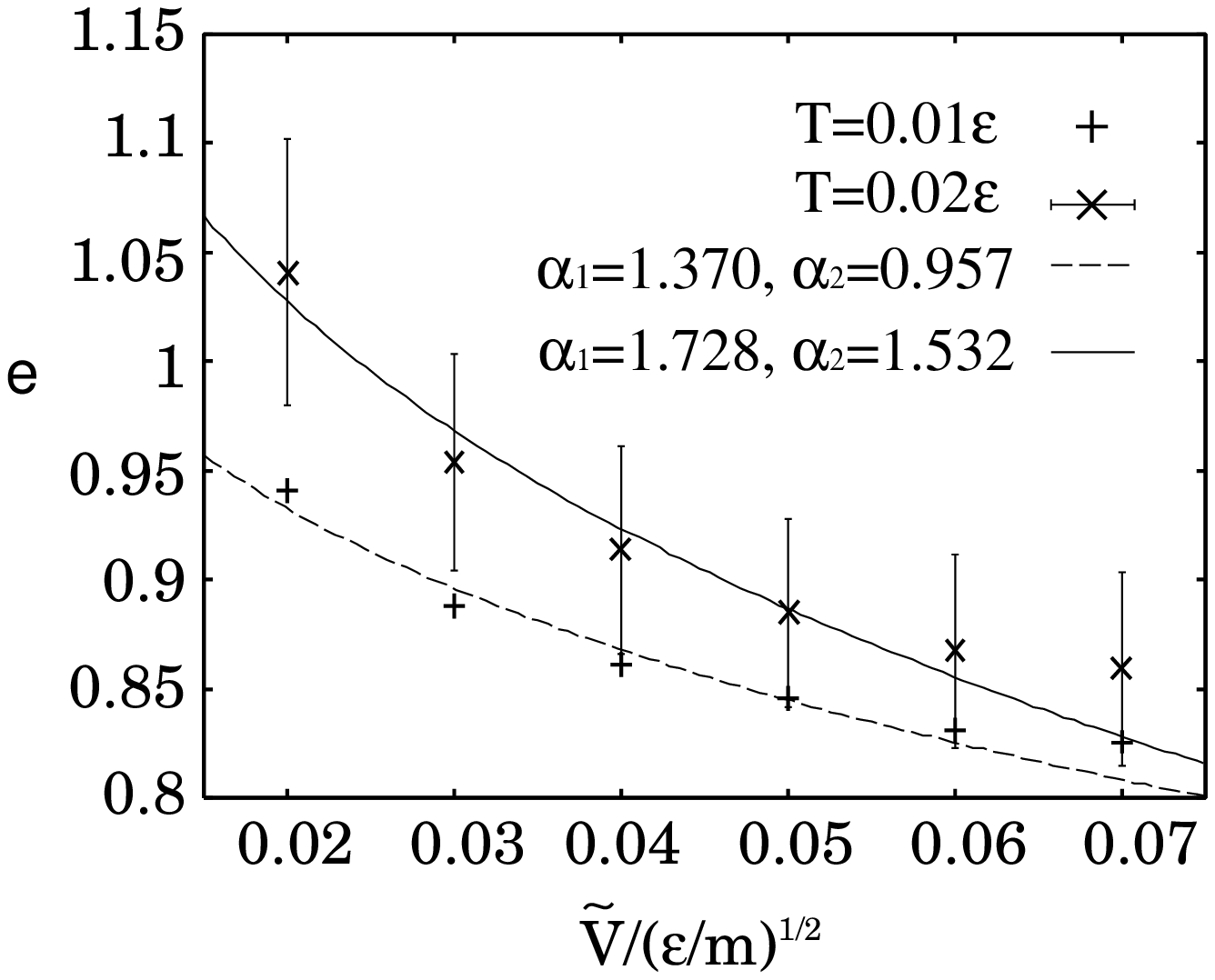}
  \caption{Relation between colliding speed and restitution coefficient. 
The solid and broken lines are results of the quasi-static theory.}
\label{fig2}
 \end{minipage}
\hspace*{3mm}
 \begin{minipage}{0.47\textwidth} 
  \includegraphics[width=0.8\textwidth]{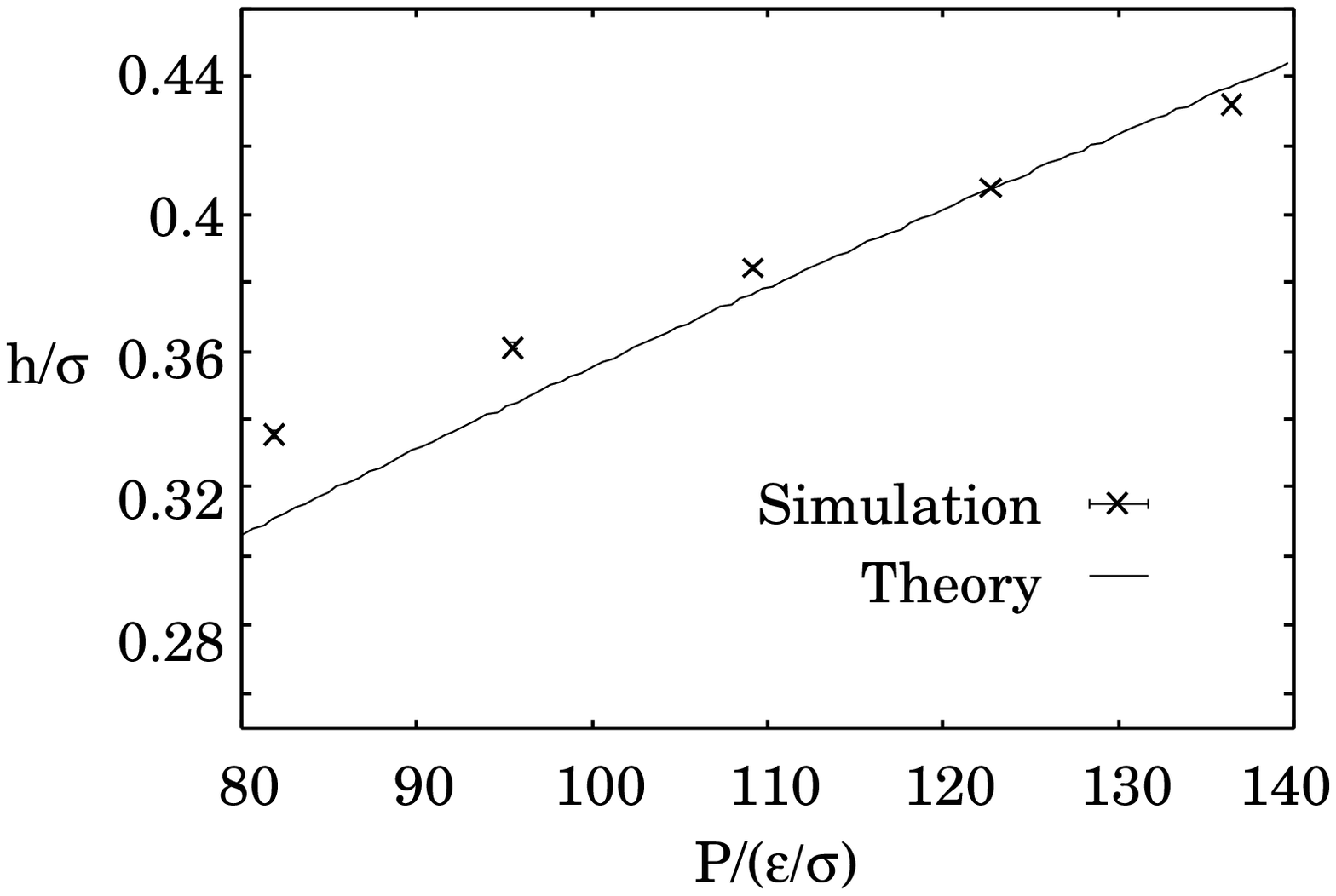}
\caption{Relation between compressive force and deformation. Cross 
points and error bars are average and standard deviation of 
$10$ numerical results. Error bars are hardly seen due to small 
standard deviation.  
Solid line is the result of Hertzian contact theory 
for two spheres pressed each other.
}
\label{fig3}
 \end{minipage}
\end{center}
\end{figure} 
Figure ~\ref{fig2} shows the relation between the relative speed of 
impact ${\tilde V}/ \sqrt{\epsilon/m}$ and the restitution coefficient $e$. 
Cross points and error bars are the average and the standard deviation 
of $100$ samples for each value on $x$-axis. 
From this result, we confirm that the restitution coefficient $e$ 
decreases with the increase of the colliding speed 
${\tilde V} / \sqrt{\epsilon/m}$. When the colliding speed is 
${\tilde V} = 0.02 \sqrt{\epsilon/m}$ at $T=0.02\epsilon$, 
the average of $e$ becomes $1.04$ 
which is slightly larger than unity.  
It is interesting that our result can be
fitted by the quasi-static theory 
of low-speed impacts $1-e\propto {\tilde V}^{1/5}
$~\cite{kuwabara,brilliantov96,morgado} 
if the restitution coefficient
at $V=0$ is replaced by a constant larger than unity. 
Indeed, the solid and the broken lines in Fig. ~\ref{fig1} are 
fitting curves of 
$e=\alpha_{1}-\alpha_{2} \left({\tilde V}/ \sqrt{\epsilon/m}\right)^{1/5}$, 
where $\alpha_{1}$ and $\alpha_{2}$ depend on material constants of colliding 
bodies. 

%
%

%
%
To remove the possibility of accidental agreement between the fitting curve
and the data, we also check the validity of Hertzian contact theory 
in our system.  
According to the theory, when the two elastic spheres are pressed 
each other, the relation between the deformation $h$ and the compressive 
force $P$ is described as $h = D P^{2/3}$ and 
$D = (3/2)^{2/3} ((1-\nu^{2})/E)^{2/3} (R/2)^{-1/3}$, 
where $\nu$ and $E$ are Poisson's ratio and Young's modulus, respectively. 
~\cite{hertz,landau}
Poisson's ratio and Young's modulus of the model 
can be calculated as follows. 



Adopting these material constants, 
we confirm that our numerical result of compression of
two clusters is well described by the Hertzian contact theory 
without any 
fitting parameters between $P=81.84\epsilon/\sigma$ 
and $P=136.4\epsilon/\sigma$ at $T=0.03\epsilon$ (Fig. ~\ref{fig3}). 
For smaller compression finite deformation seems to remain. 
For larger compression we observe too large repulsion 
because of the existence of local plastic deformations. 
Thus, we conclude that the relation 
between the impact speed and the restitution coefficient  
is characterized by the quasi-static theory of impact processes.

At last, we compare our numerical results with the fluctuation relation 
for impact phenomena which is a kind of the fluctuation theorem 
on the basis of the probability distribution 
of the macroscopic energy loss. 
In the case of impact phenomena, 
the fluctuation relation may be written as follows~\cite{hal}:
\begin{equation}\label{FR}
\exp(W(X_{0}, X_{1})/T) P(W(X_{0}, X_{1}))
=P(W({\bar X_{1}}, {\bar X_{0}})),
\end{equation}
where $X_{i} (i=0,1)$ are the macroscopic variables at 
initial and final states, respectively, 
while ${\bar X_{i}}$  are the states obtained by reversing 
all the momenta in $X_{i}$. 
$P(W(X_{0}, X_{1}))$  is the probability distribution of 
$W(X_{0},X_{1})$ which is the macroscopic energy loss during 
the transition from $X_{0}$ to $X_{1}$ defined as 
\begin{eqnarray}
 W(X_{0}, X_{1})=\sum_{i=C^{u},C^{l}}
\left[\frac{1}{2} M(V_{i}^{2}-V_{i}^{'2})
+\frac{1}{2}I(\omega_{i}^{2}-\omega_{i}^{'2})\right]\nonumber \\
+R(r)-R(r^{'}).
\end{eqnarray}
Here, $M$ is the total mass of one cluster. 
$\omega_{i}$ and $V_{i}$ are respectively 
the angular velocities and the speed of the center of mass 
for the cluster $i$ at initial state 
while $\omega_{i}^{'}$ and $V_{i}^{'}$ are respectively those at 
final state. 
$I$ is moment of inertia of the cluster. $R(r)$ represents 
the repulsion potential, where $r$ is the distance 
between the centers of mass of the two clusters.

In our simulation, we at first equilibrate the two clusters 
at $T=0.02 \epsilon$ and collide them each other with 
the initial condition ${\tilde V}=0.02 (\epsilon/m)^{1/2}$, 
 $\omega_{i}=10^{-6} (\epsilon/m)^{1/2}/\sigma$, and $r=12.02\sigma$.
Here we do not use the initial macroscopic rotations 
induced by the thermal fluctuations 
with the average $9.5 \times 10^{-7}(\epsilon/m)^{1/2}/\sigma$ 
and the standard deviation $4.9 \times 10^{-7}(\epsilon/m)^{1/2}/\sigma$ as it is. 
Instead,  for simplicity of our simulations, 
we give the initial $\omega$ whose value is approximately equal to 
$\sqrt{\langle \omega^2 \rangle}$ without taking into account
 the fluctuation of the initial rotations.
We calculate $W \equiv W(X_{0},X_{1})$ 
from the initial and the final macroscopic energy 
by assuming the final state as 
the state when the internal energy of the clusters keeps constant value 
in time. 
After the collision, 
we equilibrate the clusters at the initial temperature, 
and reverse the translational velocities to make them collide 
each other again. At the termination of the second collision, 
we calculate ${\bar W} \equiv W({\bar X_{1}},{\bar X_{0}})$. 
We obtain the probability distributions 
of $W$ and $\bar{W}$ 
from $5000$ samples. On the basis of the probability distributions, 
we investigate whether the relation (~\ref{FR}) holds in this system.

\begin{figure}[h]
\begin{center}
\includegraphics[width=.7\textwidth]{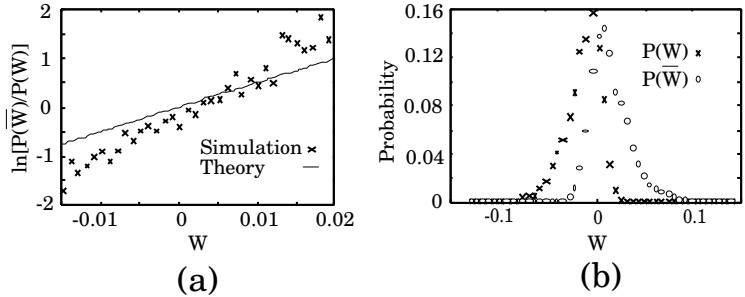}
\end{center}
\caption{(a) Numerical relation between $\ln P(\bar{W})/{P(W)}$ and $W$ for 
initial temperature $T=0.02\epsilon$.
Cross points are numerical results while 
the solid line is results from the fluctuation relation of inelastic
 impacts in eq.(~\ref{FR}). 
(b) The probability distribution of $W$ and ${\bar W}$ obtained by our 
simulation. }
\label{fig4}
\end{figure}

Figure ~\ref{fig4}(a) shows the relation 
between $W$ and 
$\ln P(\bar{W})/P(W)$. 
The solid line corresponds to $W/T$ at $T=0.02\epsilon$. 
We find that our simulation data are nearly consistent with 
the line in the range of $-0.01\epsilon < W < 0.02\epsilon$. 
Thus, the relation (~\ref{FR}) holds 
in the restricted range of $W$ in our simulations.

Here we should comment on the region in which our simulation 
results are consistent with the fluctuation relation. We show both 
$P(\bar{W})$ and $P(W)$ 
in Fig. ~\ref{fig4}(b). Both distributions are overlapped mostly 
in the range of $-0.02\epsilon<W<0.02\epsilon$. 
In the range of $|W| > 0.02 \epsilon$, 
the values of $\ln P(\bar{W})/P(W)$ 
are calculated from few samples.  
Thus, our data and the relation ~(\ref{FR}) show good agreement 
only in the range $-0.01\epsilon<W<0.01\epsilon$, 
while the discrepancy becomes large outside the region. 
Since the effect of rotations is small, 
$W$ is approximately proportional to $1-e^2$.
Thus, events with the negative $W$ almost correspond 
to ``super-elastic'' collisions. 

\section{Discussion}\label{discussion}
Let us discuss our results. 
While our model mimics impact phenomena of 
small systems subject to large thermal fluctuations, 
we should address that our model may not be adequate 
for the description of most of  realistic collisions of nanoclusters, 
where the adhesive interaction between 
clusters often prohibits the rebound in the low-speed impact. 
Thus, ``atoms'' in our model may be regarded as a coarse-grained 
units of crystalline solids and 
the dominant interaction among clusters is repulsive.
We also expect that our model 
is an idealized collision of stable fullerene~\cite{full} 
or charged clusters in which
attractive interaction between clusters is weak\cite{adh}.
As an additional remark, we should indicate that 
it may be difficult to control the colliding speed and 
the initial rotation of the cluster in actual situations 
because the macroscopic motion of one cluster is also affected 
by thermal fluctuations. 
\section{Conclusion}\label{conclusion}

In conclusion, we have performed molecular dynamics simulations 
to investigate the behaviors of colliding clusters and 
the relation between those and the fluctuation theorem. 
The results of our simulations have revealed that the relation between 
colliding speeds and the restitution coefficient can be described 
by the quasi-static theory for inelastic impacts. In addition, 
on the basis of the distribution function of macroscopic energy 
loss during collision, we have shown that our numerical results 
can be explained by the fluctuation relation for inelastic impacts.

\section*{Acknowledgements}
We would like to thank H. Tasaki for valuable comments. 
We would also like to thank M. Matsushita 
for carefully reading the manuscript and giving some advices. 
Parts of numerical computation in this work were carried out 
at Yukawa Institute Computer Facility. 
This study is partially supported by the Grant-in-Aid of 
Ministry of Education, Science and Culture, Japan (Grant No. 18540371).

%

\end{document}